# Impact of uncertainties on the Stability Lobe Diagram for vibration evaluation in milling

Diego R. Villacreses Naranjo[1][0000-0003-2971-6259], Frédéric Vignat[1][0000-0001-5316-1762] and Nicolas Béraud[1][0000-0001-5149-8644]

[1] Univ. Grenoble Alpes, CNRS, Grenoble INP, G-SCOP, 38000 Grenoble, France
diego-rodrigo.villacreses-naranjo@grenoble-inp.fr

**Abstract.** Despite being the subject of study for several years, excessive vibration persists in the machining of metal parts. In this context, the Stability Lobe Diagram (SLD) is presented as a viable tool to mitigate this problem as a function of axial depth of cut and spindle speed. However, its accurate construction is subject to the consideration of multiple parameters and models, whose application may be affected by certain inherent uncertainties. In turn, this impacts its accuracy, especially in the stability and instability regions. The present study aims to characterize these uncertainties, analyze their influence on the SLD, and propose strategies for their reduction. Ultimately, the goal is to facilitate the user's decision-making when choosing the trajectory generation parameters.

## 1    Introduction

This article addresses the crucial problem of excessive vibration in milling processes. This phenomenon has been the subject of intensive study in the last decades to understand its underlying mechanisms and develop effective mitigation strategies. However, the issue persists as one of the main constraints to productivity improvement in various machining operations [1]. Existing literature has witnessed several contributions that have outlined the factors that lead to vibration generation during machining, as well as strategies to counteract this phenomenon [2]. Previous authors have consistently identified that the interaction between the machine tool, the rotating tool, and the workpiece constitutes a scenario leading to unwanted vibrations [3] (see Fig. 1).

This dynamic and non-linear interaction represents one of the outstanding problems in the inherent complexity of machining systems [4, 5]. This complexity results in the presence of multiple modes of vibration, of which precise identification can be an arduous task. In addition, the inherent variability of machining conditions and workpiece geometry introduces an additional component of uncertainty in vibration evaluation [6].

Early on, Tobias and Fishwick [6] established the theory of regenerative chatter, which is present in mechanical part machining and poses a challenge for implementing a comprehensive model to mitigate it in the machine shop. Accurate measurement requires numerous parameters, and each machine/tool/workpiece system has its specificity. Nevertheless, their work has provided a strong foundation for future developments in this field. Notable contributions were also made by Tlusty and Polacek [3] and Merritt [7], who recognized the challenge of achieving chattering-free cutting processes but emphasized the value of a stability chart considering the self-excited chattering.

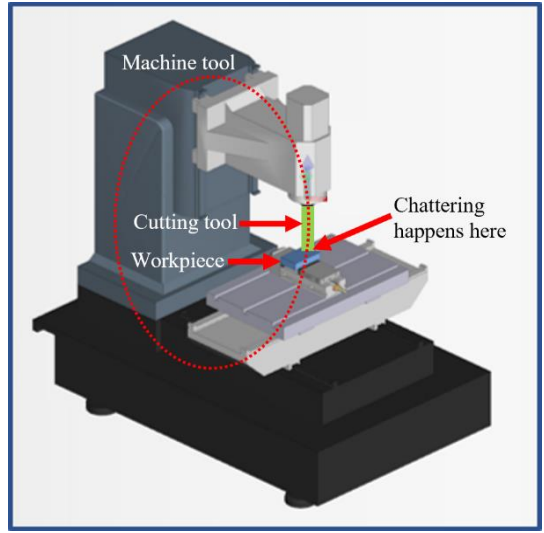

**Fig. 1.** Circuit for chattering occurrence considering a machine tool, cutting tool, and workpiece

In this framework, an important contribution has been the introduction of the delay differential equation (DDE). It is a mathematical approach used for modeling complex dynamics systems, such as the one encountered in machining processes [1]. The most general case is given by Equation 1:

$$M\ddot{x} + C\dot{x} + Kx = F \quad (1)$$

Here, the left-hand side is represented by the components of a dynamic machine-tool behavior, whereas the right-hand side is a process force [4]. The dynamic parameters of the system are measured experimentally by an experimental modal analysis (EMA) [1, 8, 9], although, analytical modeling by applying beam theory has been also used for this [9, 10]. In the case of the force, it is characterized by models fed by cutting coefficients or specific cutting forces [1]. They establish a relationship between the workpiece material properties, the interaction between the tool and the material, and the tool geometry, and are obtained via cutting tests or by orthogonal or oblique cutting models [11–13]. As with any measurement procedure, the characterization of these parameters is subject to uncertainties [14, 15].

Furthermore, different models have been developed to solve this equation. There are methods in the time domain and the frequency domain, as well as models that take into account one degree of freedom to several degrees of freedom. Different reviews of these models can be found in the literature where more details of these models can be found [16–19].

For better visualizing the system stability, the Stability Lobe Diagram (SLD) is commonly used [20]. This is a graphical tool that provides a visual representation of stability and instability regions as a function of spindle speed vs. axial depth of cut. In practical terms, stable regions in the SLD indicate combinations of parameters where the system exhibits controlled and desirable vibrational behaviors, while unstable regions indicate configurations prone to the generation of uncontrolled, and therefore, undesirable vibrations [21].

According to the researchers Munoa et al. [1] the SLD can be broken down into distinct zones (see Fig. 2). For instance, zone A, also known as the Process damping zone, highlights the importance of damping in the system, where stability is influenced by the friction generated between the tool and the irregular surface of the workpiece. In this area, it is observed that as the spindle speed decreases, the stability limit increases, resulting in a larger axial depth of cut. In contrast, zone B or intermediate indicates that the stability limit approaches the absolute limit of the system within the spindle speed range, particularly when damping values are high. Zone C, referred to as the high-speed zone, allows for an optimal combination of spindle speed and depth of cut when they lie within the lobe, where system stability is maximized. Finally, zone D, or ultra-high-speed zone, suggests that the system stability can be increased even at considerably high spindle speeds, allowing an increase in the depth of cut. However, this zone is limited by machine and power capabilities and it is a practical challenge to reach this zone in practice.

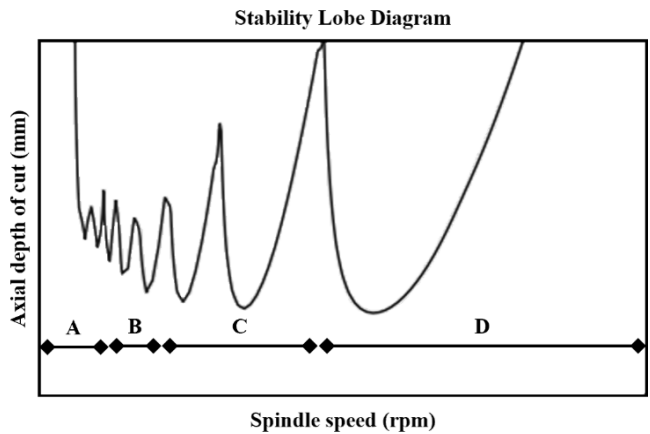

**Fig. 2.** Stability Lobe Diagram with its different zones. Zone A: Process damping zone. Zone B: Intermediate zone. Zone C: High-speed zone. Zone D: Ultra high-speed zone.

By using the SLD researchers and industry operators are allowed to effectively identify the operating conditions that maximize machining system stability [22, 23]. Additionally, from Fig. 2 it can be inferred the critical relevance of an accurate SLD construction. If the goal of the machining operator is to position in a determined zone of the diagram to seek an optimal combination of spindle speed and axial depth of cut, it is essential to ensure that the location and limits of the lobes are correct.

The focus of this study is primarily directed towards the evaluation of vibration in the context of the milling process, outlining the contributing factors and a method for its visualization to address and improve milling efficiency. The rationale of this research lies in the recognition that, despite the exhaustive studies carried out on vibration, its incidence persists.

Section 2 will portray the methodology of this study. By first recognizing the key elements that partake in the construction of the SLD, identifying the ones that possess uncertainties, and then proposing strategies to mitigate them, this study will permit to establish the basis for the future development of a decision-making tool for the selection of toolpath generation parameters during milling. In this regard, the underlying aspiration is to consolidate the information accumulated over the years and make it accessible for users of computer-aided manufacturing (CAM) software.

## 2      Methodology

The proposed methodology is based on mapping the key parameters for the SLD construction. These elements include the characteristics of the cutting tool the workpiece material and process parameters. The next step consists of identifying the uncertainties linked to the previously mapped elements. Finally, strategies on how to reduce them will be explored as a way of constructing a more precise SLD (see Fig. 3).

### 2.1     Elements mapping

The milling process is composed of a variety of elements that interact with each other to determine the success of the machining process. Within these elements are included the machines, cutting tools, and other components associated with this operation. In the context of this research, a special focus will be given to the evaluation of the vibration originating from the cutting tool, as the first phase of this analysis. Additionally, parameters regarding the workpiece and process will also be explored.

**Tool.** A careful approach is dedicated to the study and consideration of all the elements that make up the cutting tool, including both geometrical characteristics and the material of which it is composed. Such elements are integrated into a conceptual framework that aims to understand the factors that influence the generation vibrations in the milling process. These elements will serve at a later stage to construct the tool model.

**Workpiece.** The workpiece is defined primarily by its geometry and material. However, for this study, the geometry is considered a rigid block, therefore, it is implied that it will not contribute to vibration generation. Instead, attention will be paid to the workpiece material, since some coefficients can be obtained from here.

**Process parameters.** These parameters refer to the variables controlled by the user within the CAM software to define the operational details of the milling operation. Although multiple parameters are available for selection in the CAM interface, special focus is paid to those most relevant to the evaluation of vibration in the specific context of this study.

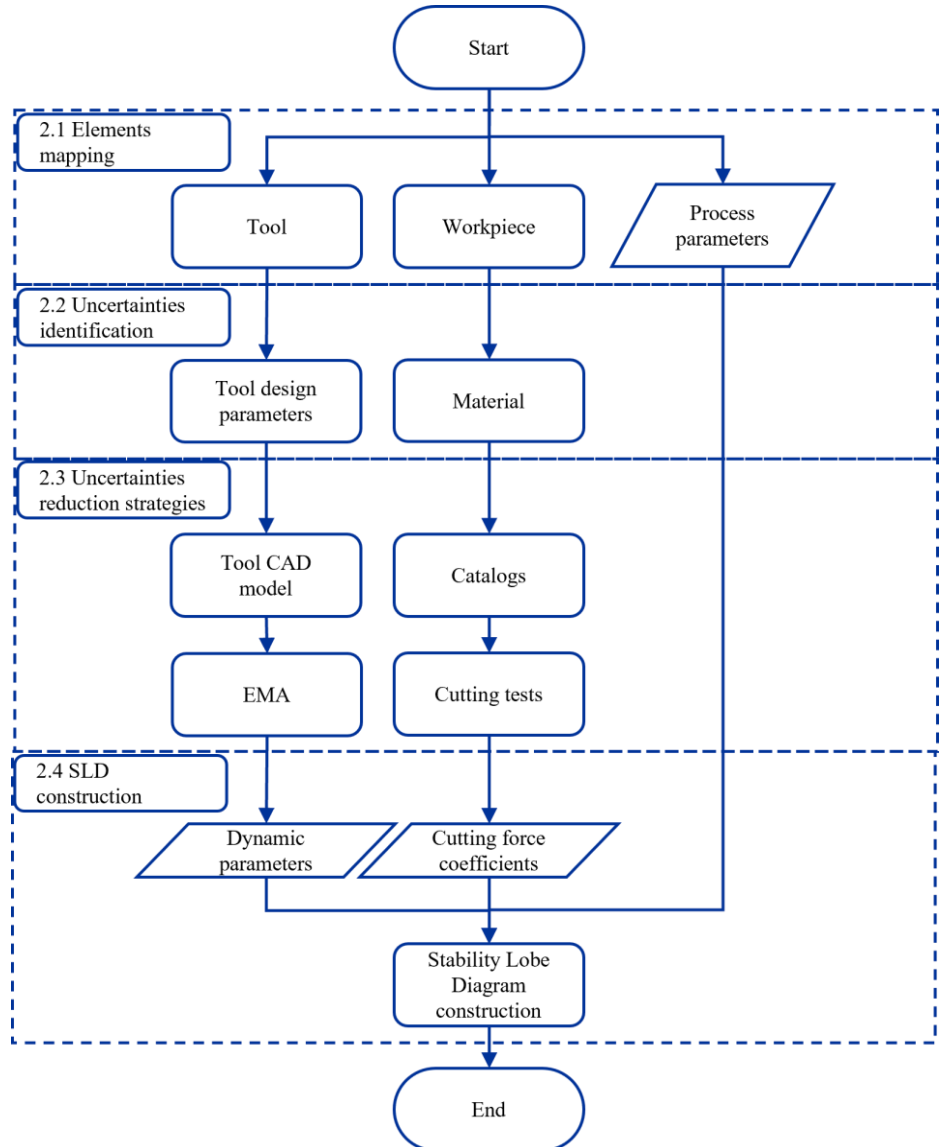

**Fig. 3.** Methodology for depicting the uncertainties for SLD construction for vibration evaluation in milling

On the other hand, Fig. 4 illustrates the results obtained during the initial stage of the proposed methodology. At this stage, a more detailed and exhaustive analysis of the literature was carried out to identify the key elements involved in the construction of the SLD. The main objective of this stage was to highlight the factors that can contribute to the uncertainties in the process. Such factors will serve as a starting point for the next stage of the study.

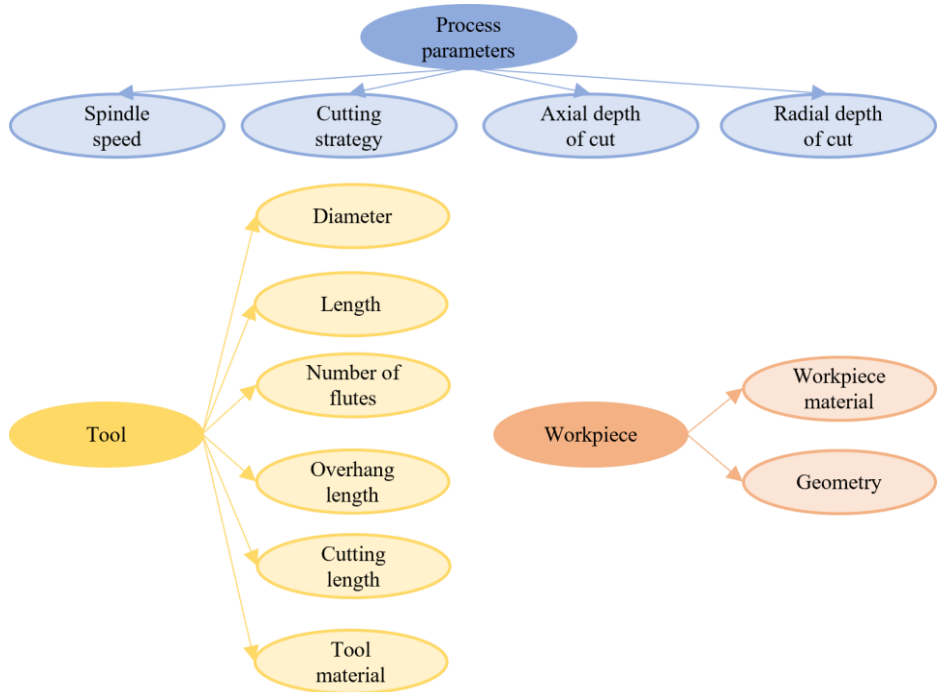

**Fig. 4.** Extended list of elements involved in the SLD construction

### 2.2 Uncertainties identification

After mapping the elements involved in the vibration evaluation, the next stage is to identify which of them have uncertainties. In the context of this study, special attention will be given to two of the three key fundamental elements in milling as established so far: the cutting tool and the workpiece.

**Tool design parameters.** As illustrated in Fig. 4, the cutting tool is defined by a variety of design parameters that affect its modal characteristics. These modal features, in turn, differ from one tool to another due to their different geometries, therefore, it is challenging to generalize the modal parameters from one tool to another.

**Workpiece material.** The workpiece is defined by both its material and its geometry. At this stage of the study, we will focus only on the workpiece material. The uncertainty here is introduced as the material composition and properties have a direct impact on the cutting force coefficients that are derived from the interaction between the tool and the workpiece interaction. Subsequently, this uncertainty can have significant implications on the accuracy and reliability of the stability diagram for vibration evaluation.

## 2.3   Uncertainties reduction strategies

Once the sources of uncertainty for the construction of the SLD have been identified, strategies for their reduction will be proposed.

**Tool CAD model.** While beam theory provides a conceptual framework for the extraction of the relevant modal parameters, it is important to note that the approximation of a solid cylindrical beam used in this theory differs greatly from the complexity and variety of geometries present in real-life cutting tools.

Although CAD models are available for some tools provided by manufacturers, not all of them can be obtained in this manner. Moreover, such models may not fully reflect the diversity of configurations and variations present in actual tools. In addition, designing custom tools from scratch can be a laborious and time-consuming process. Factors such as the number of flutes, cutting length, helix angles, and material properties exert a significant influence on the natural frequency response of the tool, as well as other relevant modal parameters.

The approach taken here is to address this complexity by modeling the tools from parameters available in CAM software. This allows the creation of case-specific tool models to facilitate the obtention of more accurate dynamic responses.

Based on the fundamentals of beam theory, the mass and stiffness matrices can be calculated from the actual tool model used in each specific case. For capturing the variations in the cross-section of the tool model, a proper discretization of the elements is required. This approach allows a more accurate representation of the dynamic response of the tool used during milling, considering its actual structure and the relevant geometrical features. However, it is important to recognize that the damping cannot be directly calculated from numerical simulations.

**Experimental Modal Analysis.** The EMA consists of fixing an accelerometer on the tooltip where a controlled impact with a hammer is performed in that region. The resulting signals are subjected to further processing, allowing the natural frequency and other relevant dynamic parameters to be obtained. However, it is essential to note that the resulting damping value derived from this experimental analysis is intrinsically linked to the specific system configuration (spindle/tool holder/tool) used during the experimentation which until this point could not be obtained from numerical simulations. Fig. 5 shows an illustration of the uncertainty reduction strategies for the dynamic parameters obtained from the tool.

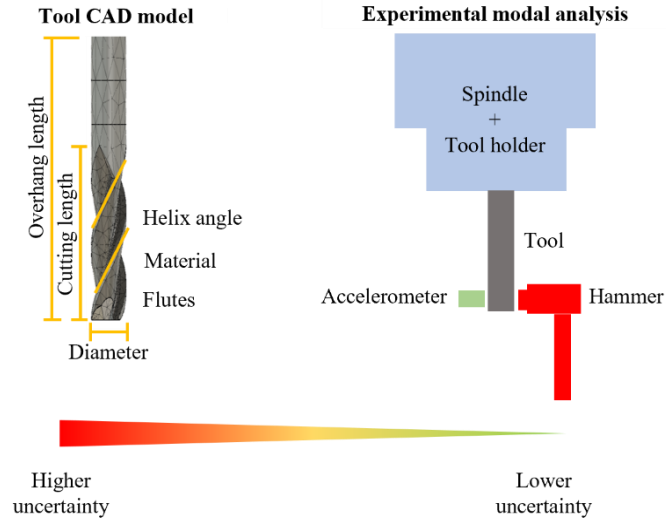

**Fig. 5.** Uncertainties reduction strategies proposal concerning the elements obtained from the tool

**Tool catalogs.** One of the main sources of uncertainty in this aspect lies in the determination of the cutting force coefficients that describe the interaction between the cutting tool and the workpiece material. A common way to obtain these coefficients is through tool catalogs provided by manufacturers. However, this approach carries a high level of uncertainty due to several reasons. For instance, these coefficients may be based on theoretical or empirical data obtained under specific cutting conditions, which in turn may differ from actual machining conditions in the workshops. Furthermore, variability in the values given by different manufacturers may introduce inconsistencies in the coefficient selection.

**Cutting tests.** To reduce the uncertainty in the determination of these coefficients, cutting tests for each specific combination of tool and workpiece material are recommended. Through this approach, it is possible to obtain direct experimental data that reflect more accurately the actual machining conditions. Experiments of such nature can be static or dynamic cutting tests in which the cutting forces that are generated during the process are recorded and analyzed. Furthermore, this approach allows experiments to be tailored to the specific characteristics of the workpiece material and its properties.

While conducting cutting experiments may be more laborious and resource- intensive than simply using tool catalogs for obtaining these coefficients, the experimental data obtained will provide a more solid and reliable basis for vibration evaluation in milling.

## 2.4 SLD construction

The SLD is a characteristic graphical representation used for chattering evaluation that provides a clear view of the regions of stability and instability as a function of the process parameters. This diagram displays intersecting lobes, delineating areas of stability (below the lobes) and instability (above the lobes). In addition, user-selected process parameters are also plotted on the SLD.

This visualization provides the user with a clear indication of the location of the chosen parameters concerning the stability and instability zones of the process. This informed analysis capability is critical for decision-making at this stage of the process, allowing the selection of a good set of parameters that maximize the efficiency and quality of the process while at the same time minimizing risks associated with vibration.

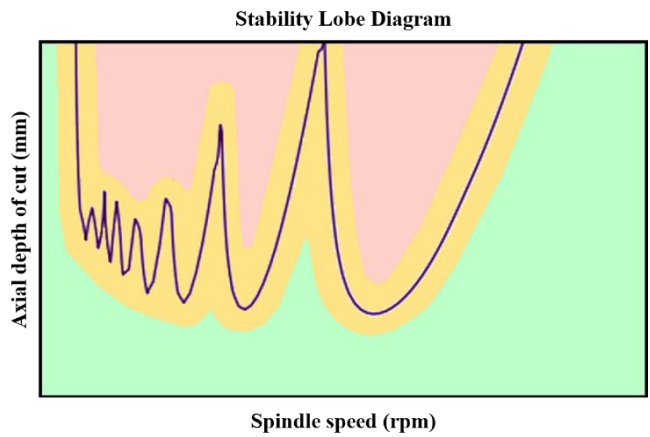

**Fig. 6.** Stability Lobe Diagram constructed considering the uncertainties inherent to their elements. The red, orange, and green zones depict the unconditionally unstable, conditionally stable (uncertain), and, unconditionally stable regions, respectively.

As shown in Fig. 6, different regions can be observed. The green zone below the lobes corresponds to the unconditionally stable region, this means, the process parameters chosen by the user will not result in vibration. The red zone above the lobes is the unconditionally unstable region, which means that the process will result in chattering. The orange zone highlights the uncertain region, therefore, the process parameters chosen will yield a conditionally stable machining process. The limits of this region are given by the uncertainty of the elements used for the SLD construction.

From a first approach in which only parameters available in the CAM software are used, to then considering additional parameters found in tool catalogs, the position of the lobes will differ, therefore, leading to the uncertainty region. Correspondingly, when a CAD model is used for the calculation of the dynamic parameters, the uncertainty region will also be affected. Until this point, the damping ratio was neglected due to the difficulty associated with obtaining its value without experimentation. The position of the lobes can more accurately be obtained by performing an EMA.

## 3 Conclusions and outlook

Of all the contributions made in this field over the last decades, chattering still represents an issue in the machining workshop. The presented methodology offers a significant contribution to the evaluation and mitigation of excessive vibrations during milling by providing an integral approach to understanding and addressing this phenomenon. Such an approach considers a wide range of factors, from tool geometry and workpiece material parameters to process parameters.

First, by identifying the key elements that influence the SLD we can better understand it for an accurate vibration evaluation. The SLD construction provides a clear visualization of the stability and instability regions, thus, allowing the operators to quickly identify parameter settings. These parameters directly influence the stability of the machining operation. By building an accurate SLD, we can identify the critical areas where vibrations are most likely to occur. Therefore, such parameters can be adjusted to minimize unwanted vibrations.

Second, while it is true that fully characterizing the damping ratio can be challenging due to the complex nature of vibrations in milling, an approximation is sought to be obtained using the damping values obtained from the tested tools. By understanding that each tool exhibits unique damping behavior, an extrapolation of this information is intended to infer damping relationships for a variety of parameter combinations. This approach allows us to consider the influence of damping on tool dynamics and ultimately improve the accuracy of our simulations and predictions in the milling process. A sensitivity analysis between several tool CAD models with varied geometries in encouraged as well for this aspect.

Ultimately, this methodology facilitates informed decision-making to optimize the selection of the parameters for generating the toolpath and minimize vibration risk during milling. By considering the variables available to the user at first instance, such as the user's knowledge, the tool catalogs, and the available CAM parameters in the software, the user can define the problem. In addition, the decision-making tool will ease the selection of the parameters required for generating the toolpath at this stage where the user can change their decision depending on the evaluation concepts taken into account for the specific machining process. This leads to increased productivity and reliability in this operation by making available the information and developments in this field to CAM users (see Fig. 7).

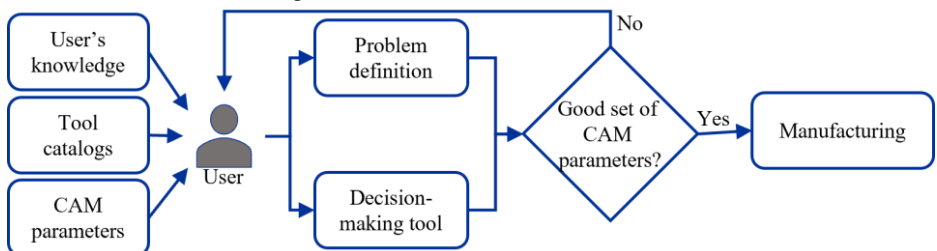

**Fig. 7.** Future works from a wholistic decision-making tool perspective

In the future, the integration of considerations for vibrations originating from the workpiece and fixturing systems will be integrated as part of this vibrations risk evaluation as well as other evaluation concepts such as tool collision, and deflection, among others. This will enable a more complete and effective management of this issue.

# 4　References